\documentstyle[epsfig,subfigure,amsmath,amssymb,float]{elsart}

\begin{document}
\bibliographystyle{elsart-num}

\def\bra#1{\mathinner{\langle{#1}|}}
\def\ket#1{\mathinner{|{#1}\rangle}}
\def\braket#1{\mathinner{\langle{#1}\rangle}}
\def\Bra#1{\left<#1\right|}                 \def\Ket#1{\left|#1\right>}
{\catcode`\|=\active
\gdef\Braket#1{\left<\mathcode`\|"8000\let|\BraVert       {#1}\right>}}
\def\BraVert{\egroup\,\mid@vertical\,\bgroup}
%\medskip

\begin{frontmatter}
\title{Feshbach resonances in a strictly two-dimensional atomic Bose gas}
\author{K.K. Rajagopal}, 
\author{P. Vignolo\corauthref{cor1}} 
and \author{M.P. Tosi}
\corauth[cor1]{Corresponding author, e-mail: {\tt vignolo@sns.it}}
\address{Classe di Scienze, Scuola Normale Superiore,\\
 Piazza dei Cavalieri 7,
  I-56126 Pisa, Italy}
  \date{}
\maketitle

\begin{abstract}
We discuss the atom-atom scattering problem across a Feshbach resonance 
in a two-dimensional dilute Bose gas at zero temperature, in the limit 
where the $s$-wave scattering length exceeds the width of the vertical 
confinement. We determine a tunable coupling-strength parameter and 
by controlling it we evaluate how the condensate wave function spreads 
out with increasing atom-atom repulsions. We also discuss the stability 
of the condensate in the magnetic-field regime where the coupling has 
become attractive.
\end{abstract}
\end{frontmatter}

\section{Introduction}

An ultracold two-dimensional (2D) Bose gas exhibits in several respects 
a different physical behaviour from its three-dimensional (3D) 
counterpart \cite{schick,popov}. In the macroscopic limit phase 
fluctuations at any finite temperature destroy the long-range order 
associated with Bose-Einstein condensation, but still allow a 
Kosterlitz-Thouless transition to a superfluid ``quasi-condensate'' 
state characterized by local phase coherence \cite{fisher}.
A true condensate is restored in a trapped 2D gas by modifications 
induced by the confinement in the density of states \cite{bagnato}, 
but the magnitude and sign of the scattering amplitude for 
atom-atom collisions can depend on external system parameters \cite{petrov}.

In fact, when the scattering length becomes comparable with the trap 
size a new two-particle scattering problem needs to be solved. 
The T-matrix for two-body collisions in vacuo at low 
momenta and energy, which is used in 3D to obtain the collisional 
coupling parameter, vanishes in the strictly 2D limit as the 
axial width of the trapping potential becomes smaller than the $s$-wave 
scattering length \cite{schick,popov}.
It is then necessary to evaluate the scattering processes between 
pairs of Bose particles by taking into account the presence of the 
surrounding bosons through a many-body T-matrix 
formalism \cite{stoof93,bijlsma97,Lee}.

The 3D $s$-wave scattering length can be tuned by means of external 
electric or magnetic fields \cite{tiesinga93,moerdijk95,vogels}. 
Within this process a resonance can appear at a particular value of 
the magnetic field and allow control of the coupling strength ranging 
from positive (repulsive) to negative (attractive) values. 
One may study new dynamical effects of collective oscillations in 
large condensates \cite{kagan} 
and the dynamics of collapsing and exploding condensates \cite{donley}.
Calculations of such Feshbach 
resonances have been given for 3D gases of $^7$Li \cite{gora},
$^{23}$Na \cite{moerdijk95}, 
$^{39}$K and $^{41}$K \cite{boesten}, $^{85}$Rb and $^{87}$Rb  \cite{vogels}, 
and $^{133}$Cs \cite{tiesinga93}.
Experimental observations have been reported for
$^{87}$Rb \cite{newbury}, $^{85}$Rb 
\cite{donley,roberts,courteille2,cornish}, $^{23}$Na \cite{inouye}, and 
$^{133}$Cs \cite{vuletic}. 
Feshbach resonances in quasi-2D atomic gases have 
been discussed by Wouters {\it et al.} \cite{wouters},
who have predicted a shift in the position of the 
resonance from the squeezing of the confinement.
 
In this work we examine the tuning of two-body collisions by means of 
a magnetic field in a strictly 2D condensate inside a harmonic trap 
at zero temperature. By modifying the coupling strength we can control 
the dimensionless physical parameter $\eta$ which is the ratio between the 
mean-field interaction energy and the harmonic-oscillator 
level spacing \cite{kagan2}.
In the case of repulsive interactions ($\eta>0$),
starting from a weakly interacting regime ($\eta\ll1$) 
and up to a strongly interacting one ($\eta\gg 1$),
we evaluate the equilibrium density profile as a function of
$\eta$ by 
solving numerically a non-linear Schr\"odinger equation by a split-step 
Crank-Nicholson discretization scheme.
In the case of attractive interactions ($\eta<0$) we can give an upper bound 
for the number of bosons which can condense without the condensate collapsing.
At variance from the 3D gas (see for instance \cite{antonius,savage}), 
we have found that in 2D this critical number does 
not depend explicitly on the strength of the harmonic confinement.

The paper is organized as follows. Section \ref{sec_T} summarizes the 
Feshbach resonance formalism and applies it to the 2D Bose gas. 
In Sec. \ref{sec2} we evaluate the equilibrium density profiles for the case 
of repulsive interactions and discuss the conditions for a stable 
condensate in the attractive coupling regime. Some concluding 
remarks are presented in Sec. \ref{conclusion}.

\section{Essential formalism}
\label{sec_T}
\subsection{Overview on scattering theory}
We start by recalling some well-known facts about a collision between 
two particles with internal degrees of freedom governed by an internal 
Hamiltonian $H^{int}$, 
\begin{equation}
H^{int}\ket{\alpha}=\epsilon_{\alpha}\ket{\alpha}
\label{one}
\end{equation}
where $\ket{\alpha}$ and $\epsilon_{\alpha}$ are the single-atom
eigenstate and its corresponding eigenvalue.
For example the internal 
Hamiltonian can describe the hyperfine interaction between nuclear 
and electronic spins and their Zeeman coupling to an external magnetic 
field $B$.
A two-body collision is described by the Hamiltonian
\begin{equation}
H=\frac{{ p}^2}{2 m_{r}}+\sum_{i=1}^{2}H_{i}^{int}+V({\bf r})
\label{three}
\end{equation}
in the center-of-mass system, with
${\bf p}$ the relative momentum and $ m_{r}$ the reduced mass
(see for instance \cite{stoof88}).
The Hamiltonian in Eq. (\ref{three}) is
the sum of a part with eigenstates
$\ket{{\alpha\beta}}$ (``channels'') tending asymptotically to a
symmetrized product of separate-atom internal states
$\ket{\alpha}$ and $\ket{\beta}$, and of a finite-range interaction
$V({\bf r})$ which couples the channels.
The energy 
associated with an eigenstate $\ket{{\alpha\beta}}$ is 
$E_{\alpha,\beta}=\varepsilon_\alpha+\varepsilon_\beta+
\hbar^2k^2_{\alpha,\beta}/2m_r$, where  $\hbar{\bf k}_{\alpha,\beta}$ 
is the relative momentum of the two incoming particles. 
The asymptotic scattering state $\Psi_{\alpha\beta}({\bf r})$ can be 
expanded on all scattering channels in the form
\begin{equation}
\Psi_{\alpha\beta}({\bf r})=\exp({i{\bf k}_{\alpha\beta}\cdot {\bf r}})
\ket{\alpha\beta}
 + \sum_{\alpha',\beta'} f_{\alpha\beta}^{\alpha'\beta'}
( {\bf k}_{\alpha\beta},{\bf k}'_{\alpha'\beta'} )
\frac{\exp({i {\bf k}'_{\alpha'\beta'}\cdot{\bf r}})}{r}
 \ket{\alpha'\beta'}\,,
\label{five}
\end{equation}
where the outgoing channel is denoted by $\ket{\alpha'\beta'}$ 
and the scattering amplitude $f_{\alpha\beta}^{\alpha'\beta'}$ is 
proportional to the T-matrix 
element $\bra{{\bf k}'_{\alpha'\beta'}}T \ket{{\bf k}_{\alpha\beta}}$.
The asymptotic magnitude of the 
momentum for the $\ket{\alpha'\beta'}$  channel is
$k_{\alpha'\beta'}=\sqrt{2 m_{r}(E_{\alpha\beta}-\epsilon_{\alpha'}-
\epsilon_{\beta'}})$, from conservation of total 
momentum and energy in the scattering process.
The channel is said to be open (closed) if the 
momentum takes positive real (imaginary) value.

The whole Hilbert space describing the spatial and spin degrees of 
freedom is now divided into a subspace P for open channels and a 
subspace Q including all closed channels \cite{feshbach}.
The state vector $\ket{\Psi}$ and the Sch\"odinger equation 
$H\ket{\Psi}=E\ket{\Psi}$ are projected onto the two subspaces by 
means of projection operators P and Q satisfying the conditions 
P + Q = 1 and PQ = 0.
The formal solution of the coupled equations obeyed by the two projected 
components is
\begin{equation}
\ket{\Psi_{Q}}=(E-H_{QQ}+i\delta)^{-1}H_{QP}\ket{\Psi_{P}}
\end{equation}
and
\begin{equation}
(E-H_{PP}-H'_{PP})\ket{\Psi_{P}}=0,
\label{elf}
\end{equation}
where $H_{PP}=PHP$ etcetera,
\begin{equation}
H'_{PP}=H_{PQ}(E-H_{QQ}+i\delta)^{-1}H_{QP}\,,
\label{twelf}
\end{equation}
and $\delta$ is a positive infinitesimal. The term
$H'_{PP}$ in Eq. (\ref{elf}) describes the Feshbach resonances. It 
represents an effective non-local, retarded interaction in the P 
subspace due to transitions to the 
Q subspace and back. The Hamiltonian $H_{PP}+H'_{PP}$
gives the effective atom-atom interaction in 
the open-channel subspace, the effects due to the existence of bound states being taken into 
account through the term $H'_{PP}$.

\subsection{Feshbach resonances in the 2D coupling}
A Feshbach resonance results when true bound states belonging to the 
closed-channel subspace match the energy of open channels: 
transient transitions may then be possible during the 
collision process \cite{moerdijk95}. 
Setting $H_{PP}=H_0 + U_1$ where $H_0$ includes the kinetic energy of 
relative motion and the internal Hamiltonian, the 
scattering matrix elements in the P subspace between 
plane-wave states with relative momenta ${\bf k}$ and ${\bf k}'$ read
\begin{equation}
\bra{{\bf k}'}T\ket{{\bf k}}\simeq
\bra{{\bf k}'}T_1\ket{{\bf k}}+\bra{\Omega_{1,+}{\bf k}'}H'_{PP}
\ket{\Omega_{1,+}{\bf k}}\,.
\label{intermedia}
\end{equation}
Here $T_1=U_1+U_1G_0T_1$ represents the scattering process if Q subspace
is neglected and $\Omega_{1,+}=(1-G_0U_1)^{-1}$ is the wave operator
due to the interaction potential $U_1$, with $G_0=(E-H_0)^{-1}$ 
\cite{pethick}.
In Eq. (\ref{intermedia}) the effect of closed channels has been 
included at first order in $H'_{PP}$.
In the limit of zero relative velocity Eq. (\ref{intermedia}) leads to 
an expression for the coupling strength,
\begin{equation}
g=\tilde g+\sum_{n}\frac{|\bra{\psi_{n}}
H_{QP}\ket{\psi_{0}}|^2}{E_{th}-E_{n}}
\label{olala}
\end{equation}
where the sum is over all the states $\ket{\psi_n}$ in Q subspace
and we have set $\ket{\Omega_{1,+}{\bf k}}\simeq\ket{\Omega_{1,+}{\bf k}'}
\equiv\ket{\psi_0}$. Here, $\tilde g$ is the coupling strength estimated 
in the absence of closed channels
and $E_{th}$ is the threshold energy for the state $|\psi_0\rangle$
at vanishing kinetic energy.
Finally, when the threshold energy is close to the energy $E_{res}$ of 
one specific bound state $\ket{\psi_{res}}$, this contribution will be 
dominant and the contributions from all other states may be included in   
$\tilde g$ through an effective 
non-resonant scattering length $a_{nr}$. 
For a strictly 2D condensate we get
\begin{equation}
g = \frac{4\pi\hbar^2/m}{\ln |4\hbar/\mu m a^2_{nr}|}+ 
\frac{|\bra{\psi_{res}}H_{QP}\ket{\psi_{0}}|^2}{E_{th}-E_{res}} \,,
\label{7teen}
\end{equation}
where $\mu$ is the chemical potential of the condensate.

Equation (\ref{7teen}) is our main result. 
The non-resonant first term has been calculated from the 
two-body T-matrix for a binary collision occurring at low momenta 
and energy in the presence of a condensate, which fixes the energy 
of each colliding atom at the chemical potential \cite{morgan,rajagopal}.
The resonant second term can be tuned by exploiting the dependence of 
the energy denominator on external parameters and in particular on the 
strength of the applied magnetic field. Explicitly, 
if the energy denominator vanishes at a value $B_0$ of the field, 
then by expansion around this value we have
\begin{equation}
E_{th}-E_{res}\approx ({\frak{m}}_{res}-{\frak{m}}_{\alpha}
-{\frak{m}}_{\beta})(B-B_{0})
\label{8teen}
\end{equation}
where ${\frak{m}}_{\alpha,\beta}=-{\partial\epsilon_{\alpha,\beta}}/
{\partial B}$
are the magnetic moments of the two colliding atoms in the open channel
and ${\frak{m}}_{res}=-{\partial E_{res}}/{\partial B}$ is the magnetic
moment of the molecular bound state. 
Equation (\ref{7teen}) then yields
\begin{equation}
g = \frac{4\pi\hbar^2}{m} 
\left( \frac{1}{\ln |4\hbar/\mu m a^2_{nr}|}+ 
\frac{\Delta B}{B-B_{0}}\right)\,,
\label{9teen}
\end{equation}
where the width parameter $\Delta B$ is defined as
\begin{equation}
 \Delta B =\frac{m}{4\pi\hbar^2} 
\frac{|\bra{\psi_{res}}H_{QP}\ket{\psi_{0}}|^2}
    {\frak{m}_{res}-\frak{m}_{\alpha}-\frak{m}_{\beta}}\,.
\label{wenty}
\end{equation}
Higher-order terms in $H'_{PP}$  will lead to a broadening of the resonance. 
However, the width of resonant states close to the threshold energy 
in the open channel is usually small because of the 
low density of states, and Feshbach resonances are usually very 
sharp \cite{pethick}.

In the next Section we will analyze the consequences of tuning 
the coupling strength on the 
equilibrium state of a 2D condensate in a harmonic trap.
 
\section{Equilibrium properties as functions of the coupling}
\label{sec2}

We consider a Bose gas with a fixed number $N$ of particles in a 2D harmonic 
potential $V_{ext}(r)=m\omega^2 r^2/2$.  
At zero temperature the condensate wave function 
$\Psi(r)$ is described by the mean-field 
Gross-Pitaevskii equation    
\begin{equation}
\left[-\frac{\hbar^2}{2m}\frac{1}{r}\frac{\partial}{\partial r}
\left(r \frac{\partial}{\partial r}\right)+V_{ext}(r)+g n(r)\right]
\Psi(r)=\mu \Psi(r) \,,
\label{wentydue}
\end{equation}
where $n(r)=|\Psi(r)|^2$ is the density of the condensate.
We introduce the dimensionless coordinate $x=r/a_{ho}$,
where $a_{ho}=(\hbar/m\omega)^{1/2}$ is the harmonic oscillator length, 
and the dimensionless chemical potential $\mu'=2 \mu/\hbar
\omega$.  
Thus Eq. (\ref{wentydue}) can be written in rescaled form as
\begin{equation}
\left[-\frac{d^2}{d x^2}- \frac{1}{x}\frac{d}{d x}\right]\Psi(x)
 + x^2\Psi(x) + 2\eta \left|\frac{\Psi(x)}{\Psi(0)}\right|^2\Psi(x)=
\mu' \Psi(x) \,,
\label{wentytre}
\end{equation}
where $\eta = g\,|\Psi(0)|^2/\hbar\omega$ 
is a dimensionless coupling parameter given by the ratio between the 
mean-field interaction energy per particle and the oscillator level spacing.
The normalization condition on the wave-function is
\begin{equation}
N = a_{ho}^2\int_{0}^{\infty}x\,dx \,|\Psi(x)|^2\,.
\label{wentyfive}
\end{equation}
Equations (\ref{wentytre}) and (\ref{wentyfive}) 
need to be solved self-consistently. 

\subsection{Repulsive coupling ($\eta > 0$)}
\label{code-bil}
\begin{figure}
\centering{
\epsfig{file=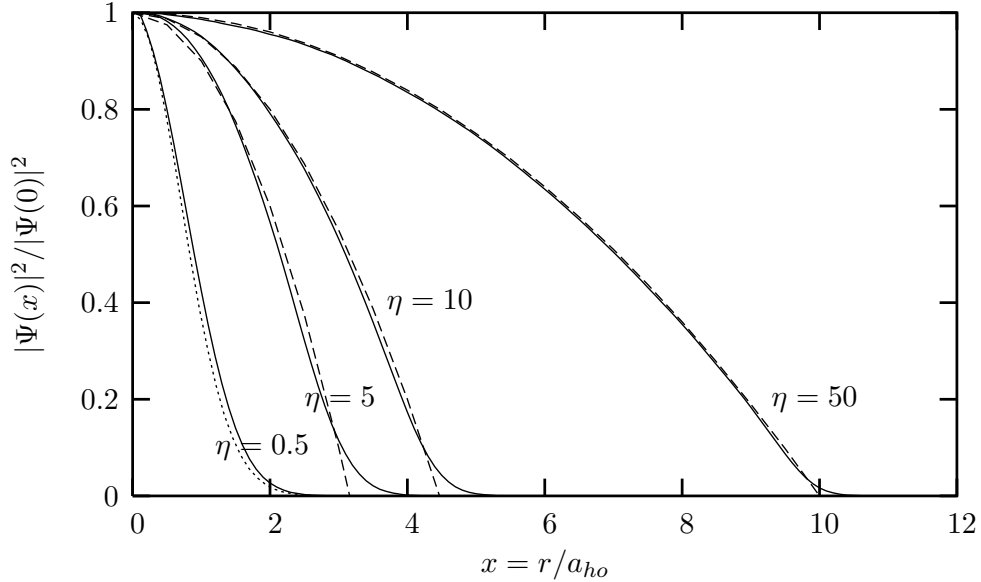,width=1.0\linewidth}}
\caption{Scaled density profile $|\Psi(x)|^2/|\Psi(0)|^2$ as a function
of the dimensionless radial distance $x=r/a_{ho}$ 
for various values of the coupling parameter
$\eta$. Dashed curves are Thomas-Fermi profiles and the dotted 
curve is the free-gas profile.}
\label{fig1}
\end{figure}

In the case of repulsive interactions Eq. (\ref{wentytre}) always admits 
a solution: the condensate exists and is stable in the 2D harmonic trap. 
Of course, far away from the resonance the dimensionless 
coupling parameter $\eta$ is primarily determined by the non-resonant 
scattering length.
On approaching the resonance by increasing $B$ towards $B_0$, 
the coupling parameter becomes very 
small for $B\sim B_0-\Delta B/(\ln |4\hbar/\mu m a^2_{nr}|)$: 
the non-linear term in Eq. (\ref{wentytre}) becomes small and the 
condensate wave function is close to a Gaussian profile. 
On further increasing $B$ the gas enters 
the strong-coupling regime and in the limit $\eta \gg 1$
the kinetic energy term in Eq. (\ref{wentytre}) and the 
discrete structure of the trap levels become irrelevant. 
The profile then approaches a Thomas-Fermi form \cite{goldman,huse}.
The shape of the condensate density profile, obtained by the numerical 
solution of Eq. (\ref{wentytre}) at 
several values of the coupling parameter $\eta$, is depicted in 
Fig. \ref{fig1}. The Figure illustrates comparisons with the free-gas and 
Thomas-Fermi profiles, according to the discussion given just above.

\subsection{Attractive coupling ($\eta < 0$)}
On going through the resonance the Thomas-Fermi-like density profile will suddenly shrink and 
collapse. Indeed, in the strong-coupling attractive regime the kinetic energy does not suffice to 
compensate for the negative mean-field energy term and Eq. (\ref{wentytre})
does not admit physical 
solutions. A condensate with a limited number of particles can only exist in the weakly attractive 
regime.
An upper bound for the critical number $N_c$ of bosons in a condensate with weakly attractive 
interactions can be estimated by comparing the kinetic energy and the mean-field energy at the 
centre of the trap. This leads to the inequality
\begin{equation}
1+2\eta\ge0
\label{eq1-2}
\end{equation}
for condensate stability, where in the calculation of the kinetic energy we have used the free-gas 
wave function. Using the expression of $\eta$ we find from Eq. (\ref{eq1-2})
\begin{equation}
N_c\simeq\left|\frac{1}{\ln |4\hbar/\mu m a^2_{nr}|}+ 
\frac{\Delta B}{B-B_{0}}\right|^{-1}.
\end{equation}
At variance from the trapped 3D gas \cite{antonius},
the critical number $N_c$ does not depend on the 
harmonic oscillator length, the dependence being cancelled by the fact 
that in 2D the mean-field 
energy scales as $\approx a_{ho}^{-2}$ like the kinetic energy.
Let us remark that the condensate in the case of attractive coupling 
can be metastable even if the number of condensed bosons is lower than $N_c$. 
A three-body loss term should be included in Eq. 
(\ref{wentydue}) in order to calculate the decay time and to describe 
the time evolution of such a condensate, 
as was done for the 3D Bose gas \cite{savage,saito,sadhan} 
and very recently for a 3D boson-fermion mixture \cite{adikari}. 
An evaluation of three-body losses in the 2D Bose gas and of the time 
evolution of a 2D condensate in the weakly attractive coupling regime 
may be considered in the future.

\section{Concluding remarks}
\label{conclusion}
In summary, we have applied the formalism already developed to 
treat Feshbach resonances in 3D boson 
gases to the case of a 2D Bose gas at zero temperature. 
One of the main results is that an 
external magnetic field could be used to drive the coupling 
from repulsive to attractive even in a 
pancake geometry as the scattering length becomes larger than the 
axial width of the condensed cloud.
For the case of repulsive interactions we have evaluated the particle 
density profiles at varying coupling strength, demonstrating the 
approach to a Thomas-Fermi profile followed by collapse. 
In the opposite regime of attractive coupling we have given an estimate 
for the critical number of bosons that can form a stable condensate. 
We have found that this critical number is inversely 
proportional to the absolute magnitude of the coupling, 
independently of the strength of the in-plane confinement.

\ack
The code used in the numerical calculations reported in 
Sec. \ref{code-bil} was originally developed 
by Professor B. Tanatar and his coworkers. 
We also acknowledge support from Scuola Normale 
Superiore di Pisa through an Advanced Research Initiative. 
One of us (MPT) thanks Professor V. E. Kravtsov and the 
Condensed Matter Theory group of the Abdus Salam ICTP for their 
hospitality during the final stages of this work.

%\bibliography{tmatrix}

\end{document}